\documentclass[aps,eqsecnum,superscriptaddress,nofootinbib,%
amsmath,amssymb]{revtex4}
\newcommand{\half}{ \frac{1}{2} }
\usepackage{bm}
\bmdefine{\bma}{ \bm{a} }
\bmdefine{\bmb}{ \bm{b} }
\bmdefine{\bmc}{ \bm{c} }
\bmdefine{\bmd}{ \bm{d} }
\bmdefine{\bme}{ \bm{e} }
\bmdefine{\bmf}{ \bm{f} }
\bmdefine{\bmg}{ \bm{g} }
\bmdefine{\bmh}{ \bm{h} }
\bmdefine{\bmi}{ \bm{i} }
\bmdefine{\bmj}{ \bm{j} }
\bmdefine{\bmk}{ \bm{k} }
\bmdefine{\bml}{ \bm{l} }
\bmdefine{\bmm}{ \bm{m} }
\bmdefine{\bmn}{ \bm{n} }
\bmdefine{\bmo}{ \bm{o} }
\bmdefine{\bmp}{ \bm{p} }
\bmdefine{\bmq}{ \bm{q} }
\bmdefine{\bmr}{ \bm{r} }
\bmdefine{\bms}{ \bm{s} }
\bmdefine{\bmt}{ \bm{t} }
\bmdefine{\bmu}{ \bm{u} }
\bmdefine{\bmv}{ \bm{v} }
\bmdefine{\bmw}{ \bm{w} }
\bmdefine{\bmx}{ \bm{x} }
\bmdefine{\bmy}{ \bm{y} }
\bmdefine{\bmz}{ \bm{z} }
\bmdefine{\bmA}{ \bm{A} }
\bmdefine{\bmB}{ \bm{B} }
\bmdefine{\bmC}{ \bm{C} }
\bmdefine{\bmD}{ \bm{D} }
\bmdefine{\bmE}{ \bm{E} }
\bmdefine{\bmF}{ \bm{F} }
\bmdefine{\bmG}{ \bm{G} }
\bmdefine{\bmH}{ \bm{H} }
\bmdefine{\bmI}{ \bm{I} }
\bmdefine{\bmJ}{ \bm{J} }
\bmdefine{\bmK}{ \bm{K} }
\bmdefine{\bmL}{ \bm{L} }
\bmdefine{\bmM}{ \bm{M} }
\bmdefine{\bmN}{ \bm{N} }
\bmdefine{\bmO}{ \bm{O} }
\bmdefine{\bmP}{ \bm{P} }
\bmdefine{\bmQ}{ \bm{Q} }
\bmdefine{\bmR}{ \bm{R} }
\bmdefine{\bmS}{ \bm{S} }
\bmdefine{\bmT}{ \bm{T} }
\bmdefine{\bmU}{ \bm{U} }
\bmdefine{\bmV}{ \bm{V} }
\bmdefine{\bmW}{ \bm{W} }
\bmdefine{\bmX}{ \bm{X} }
\bmdefine{\bmY}{ \bm{Y} }
\bmdefine{\bmZ}{ \bm{Z} }
\bmdefine{\bmalpha}{ \bm{\alpha} }
\bmdefine{\bmbeta}{ \bm{\beta} }
\bmdefine{\bmgamma}{ \bm{\gamma} }
\bmdefine{\bmdelta}{ \bm{\delta} }
\bmdefine{\bmeps}{ \bm{\epsilon} }
\bmdefine{\bmzeta}{ \bm{\zeta} }
\bmdefine{\bmeta}{ \bm{\eta} }
\bmdefine{\bmtheta}{ \bm{\theta} }
\bmdefine{\bmiota}{ \bm{\iota} }
\bmdefine{\bmkappa}{ \bm{\kappa} }
\bmdefine{\bmlambda}{ \bm{\lambda} }
\bmdefine{\bmmu}{ \bm{\mu} }
\bmdefine{\bmnu}{ \bm{\nu} }
\bmdefine{\bmxi}{ \bm{\xi} }
\bmdefine{\bmomicron}{ \bm{o} }
\bmdefine{\bmpi}{ \bm{\pi} }
\bmdefine{\bmrho}{ \bm{\rho} }
\bmdefine{\bmsig}{ \bm{\sigma} }
\bmdefine{\bmtau}{ \bm{\tau} }
\bmdefine{\bmupsilon}{ \bm{\upsilon} }
\bmdefine{\bmphi}{ \bm{\phi} }
\bmdefine{\bmchi}{ \bm{\chi} }
\bmdefine{\bmpsi}{ \bm{\psi} }
\bmdefine{\bmomega}{ \bm{\omega} }
\bmdefine{\bmDelta}{ \bm{\Delta} }
\bmdefine{\bmGamma}{ \bm{\Gamma} }
\bmdefine{\bmLambda}{ \bm{\Lambda} }
\bmdefine{\bmnabla}{ \bm{\nabla} }

\newcommand{\calB}{ \mathcal{B} }

\newcommand{\calE}{ \mathcal{E} }

\newcommand{\calJ}{ \mathcal{J} }

\newcommand{\tilD}{ \Tilde{D} }
\newcommand{\vecx}{ \Vec{x} }
\newcommand{\hJ}{ \Hat{J} }
\begin{document}
\title{Hall effect in Noncommutative spaces}
\author{Akira Kokado}
\email{kokado@kobe-kiu.ac.jp}
\affiliation{Kobe International University, Kobe 658-0032, Japan}
\author{Takashi Okamura}
\email{okamura@kcs.kwansei.ac.jp}
\affiliation{Department of Physics, Kwansei Gakuin University,
Sanda 669-1337, Japan}
\author{Takesi Saito}
\email{tsaito@k7.dion.ne.jp}
\affiliation{Department of Physics, Kwansei Gakuin University,
Sanda 669-1337, Japan}
\date{\today}
\begin{abstract}
In order to investigate whether space coordinates are
intrinsically noncommutative, 
we make use of the Hall effect on the two-dimensional plane.
We calculate the Hall conductivity in such a way that
the noncommutative $U(1)$ gauge invariance is manifest.
We find that the noncommutativity parameter $\theta$
does not appear in the Hall conductivity itself,
but the particle number density of electron depends on $\theta$.
We point out that the peak of particle number density differs from
that of the charge density.
\end{abstract}
\maketitle
\section{Introduction}
If gravity and quantum mechanics are unified in a small scale,
the manifold picture of spacetime has no operational meaning,
and then any spacetime uncertainty is expected below the Planck scale
\cite{ref:DoplicherEtAL,ref:Yoneya}.
The spacetime uncertainty is naturally brought by
noncommutative product between spacetime coordinates,
\begin{align}
  [~x^\mu~,~x^\nu~]_\star := x^\mu \star x^\nu - x^\nu \star x^\mu
   = i~\theta^{\mu\nu}~,
\label{eq:NC-coordinate}
\end{align}
where $\theta^{\mu\nu}$ is the real anti-symmetric parameter and
the Moyal product $\star$ which is associative, but not commutative
is defined by
\begin{align}
  f(x) \star g(x) := \exp\left( + \frac{i}{2}~\theta^{\mu\nu} \,
       \partial^\xi_\mu \, \partial^\eta_\nu \right)~
         f( x + \xi ) \, g( x + \eta )~\Big\vert_{\xi=\eta=0}~.
\label{eq:def-star-product}
\end{align}
$[~,~]_\star$ is called the Moyal bracket.

Recently, remotivated by string theory arguments
\cite{ref:SeibergWitten-NC}
noncommutative (NC) spacetime have been drawn much attention in
field theories
\cite{ref:Filk,ref:Schomerus,ref:MinwallaEtAL,ref:Hayakawa,%
ref:MatusisEtAL,ref:DouglasNekrasov}
as well as their phenomenological implications
\cite{ref:MocioiuEtAL,ref:CarrollEtAL,ref:HewettEtAL,ref:CarlsonEtALI,%
ref:AnisimovEtAL,ref:CarlsonEtALII}.
The NC spacetime can be realized
by replacing usual commutative product with
the noncommutative Moyal star product Eq.(\ref{eq:def-star-product}),
so that, by replacing the usual product with the Moyal star product,
we can construct the action for field theories
on NC spacetime from that on commutative spacetime.
One of the most interesting things of this procedure is
that for gauge theories on the NC spacetime,
even the $U(1)$ gauge group has non-Abelian like characters
such as self-interactions.

There are many papers concerning
the Hall effect on two-dimensional NC space
\cite{ref:DuvalHorvathy,ref:DayiJellal,ref:KokadoEtAL}
where $\theta^{0i}=0$ and $\theta^{ij} \ne 0$ $(i,~j=1, 2)$
in Eq.(\ref{eq:NC-coordinate}).
However, the results seem to be divergent, 
some show deviations
\cite{ref:DayiJellal,ref:KokadoEtAL}
and others no deviations from the usual commutative theory
\cite{ref:DuvalHorvathy}. 
This may come from the fact that though they have discussed
the Hall effect based on NC quantum mechanics,
but the NC U(1) gauge invariance was not so clear in their papers.
In this paper we would like to reinvestigate the Hall effect,
especially preserving the NC $U(1)$ gauge invariance of our system.

In Sec.\ref{subsec:general}, in order to calculate
the Hall conductivity by classical noncommutative field theory (NCFT),
we briefly review the NC $U(1)$ gauge theory.
In Sec.\ref{subsec:model}, we find that
the Hall conductivity on NC space does not depend on
the NC parameter $\theta$.
The final section is devoted to concluding remarks.
There, we point out that the center of the charge density
is different from that of the number density.
\section{Hall effect on noncommutative space}
\label{sec:Hall}
\subsection{NC Schr\"odinger field coupled to NC $U(1)$ field}
\label{subsec:general}
We consider the NC Schr\"odinger field $\psi$ coupled to
the NC $U(1)$ gauge field $A_\mu$ on noncommutative spacetime.

The action in interest is given by
\begin{align}
  S = - \frac{1}{4}~\int d^n x~F_{\mu\nu} \star F^{\mu\nu}
   + \int d^n x~\psi^\dagger \star \left( i D_0
                    + \frac{1}{2m}~D_i \star D^i \right) \star \psi~,
\label{eq:def-S}
\end{align}
where the star product $\star$ is defined by
Eq.(\ref{eq:def-star-product}).

The covariant derivative $D_\mu$ of the NC $U(1)$ gauge acts on 
fundamental representation $f(x)$ and on adjoint one $G(x)$, such that
\begin{align}
  & D_\mu \star f(x) := \partial_\mu f(x) - i g~A_\mu(x) \star f(x)~,
\label{eq:def-D-on-fundamental} \\
  & D_\mu \star G(x) := \partial_\mu G(x)
         - i g~[~A_\mu(x)~,~G(x)~]_\star =: \tilD_\mu \star G(x)~,
\label{eq:def-D-on-adjoint}
\end{align}
respectively.
Where $[~,~]_\star$ is the Moyal bracket.
In order to manifest the covariant derivative
on {\it adjoint representation} field, we often use $\tilD_\mu$.

The field strength $F_{\mu\nu}$ are defined as
\begin{align}
  F_{\mu\nu} := \frac{i}{g}~[~D_\mu~,~D_\nu~]_\star
   = \partial_\mu A_\nu(x) - \partial_\nu A_\mu(x) 
       - i g [~A_\mu~,~A_\nu~]_\star~.
\label{eq:def-F}
\end{align}

The above action is invariant under the NC $U(1)$ gauge transformation
\begin{align}
  \psi(x) & \mapsto \psi'(x) = U(x) \star \psi(x)~,
\label{eq:gauge-tr-psi} \\
  A_\mu(x) & \mapsto A_\mu'(x)
    = U(x) \star A_\mu(x) \star U^\dagger(x)
        + \frac{i}{g}~U(x) \star \partial_\mu U^\dagger(x)~,
\label{eq:gauge-tr-A}
\end{align}
where $U(x)$ is a star unitary function
$U(x) \star U^\dagger(x) = U^\dagger(x) \star U(x) =1$.

From the action (\ref{eq:def-S}), the equations of motion follow
\begin{align}
  & \tilD_\nu \star F^{\mu\nu} = J^\mu~,
\label{eq:EOM-A} \\
  & 0 = \left( i D_0 + \frac{1}{2m}~D_i \star D^i \right)
                  \star \psi~,
\label{eq:EOM-psi}
\end{align}
where $J^\mu$ is the NC $U(1)$ charge current density defined by
\begin{align}
  J^\mu := 
  \begin{cases}
    g~\psi \star \psi^\dagger & \mbox{~for~~} \mu = 0 \vspace{0.2cm} \\
   \displaystyle{
    \frac{g}{2\, m\, i}~\left[~
                ( D^\mu \star \psi ) \star \psi^\dagger
              - \psi \star ( D^\mu \star \psi )^\dagger~\right] }
   = \frac{g}{m}~
      \Im \left[~( D^\mu \star \psi ) \star \psi^\dagger~\right]
                            & \mbox{~for~~} \mu \ne 0
  \end{cases}~.
\label{eq:def-J}
\end{align}

It is apparent from the expression (\ref{eq:def-J}) that
the NC $U(1)$ charge current density $J^\mu$ transforms
adjoint-likely under the NC $U(1)$ gauge transformations, {\i.e.},
$\displaystyle{J^\mu \mapsto J'^\mu = U \star J^\mu \star U^\dagger }$.
\subsection{Hall conductivity}
\label{subsec:model}
Hereafter we confine ourselves into the case of
two-dimensional noncommutative space, which is characterized by
the Moyal star bracket
\begin{align}
  [~x^i~,~x^j~]_\star = i~\theta^{ij} = i~\theta~\epsilon^{ij}
\hspace{1cm} ( i,~j = 1, 2 )~,
\label{eq:fund-star-commutator}
\end{align}
where
\begin{align}
  & \epsilon_{ij} := \begin{pmatrix} 0 & 1 \\ -1 & 0 \end{pmatrix}~,
&
  & \epsilon^{ij} := \begin{pmatrix} 0 & 1 \\ -1 & 0 \end{pmatrix}~,
&
  & \epsilon^{ik} \epsilon_{jk} = \epsilon_{jk} \epsilon^{ik}
           = \delta^i{}_j~.
\label{eq:def-eps}
\end{align}

We take a gauge
\begin{align}
  & A_0 = \calE~x^1~, \hspace{2cm} A_1 = 0~, &
  & A_2 = \calB~x^1~.
\label{eq:def-gaugeA}
\end{align}
Then field strengths become
\begin{align}
  & E_1 := F_{10} = \calE~, & 
  & B_3 := F_{12} = \calB~,
\label{eq:field-strength}
\end{align}
and the others vanish.
In this case, the covariant derivatives in Eq.(\ref{eq:EOM-psi})
take the following forms
\begin{align}
  & D_0 \star \psi = \partial_0~\psi - i~g\calE~x^1 \star \psi~,
\label{eq:covariant-deriZero} \\
  & D_1 \star \psi = \partial_1~\psi~,
\label{eq:covariant-deriOne} \\
  & D_2 \star \psi = \partial_2~\psi - i~g\calB~x^1 \star \psi~.
\label{eq:covariant-deriTwo}
\end{align}
Since these quantities have no explicit dependence of $t$ and $x^2$,
we seek a solution such as
\begin{equation}
  \psi = e^{-i \omega t}~\phi(x^1) \star e^{i p_2 x^2}~.
\label{eq:Separable}
\end{equation}
Substituting Eq.(\ref{eq:Separable}) into
Eqs.(\ref{eq:covariant-deriZero})-(\ref{eq:covariant-deriTwo}),
we find that Eq.(\ref{eq:EOM-psi}) is turned out to be of the form
\begin{align}
  & \left[~\omega + g \calE x^1 - \frac{1}{2m}~
           \left\{~-\partial_1^2 + ( g\calB x^1 - p_2 )^2~\right\}~
             \right] \phi(x^1) \star e^{i p_2 x^2} = 0~.
\label{eq:reduced-EOM-pre}
\end{align}
Since the inverse of $\exp( i~p_2\, x^2 )$ with respect to
the star product is given by $\exp( - i~p_2\, x^2 )$,
the last exponential factor may be dropped out according to
the star product associative law.
Then we have an equation similar to the Schr\"odinger equation
of a one-dimensional harmonic oscillator,
\begin{align}
  E~\phi = \left( \frac{-1}{2m}~\partial_X^2
         + \frac{m \omega_c^2}{2} X^2 \right) \phi~,
\label{eq:reduced-EOMII}
\end{align}
where $\displaystyle{ \omega_c := g\calB/m }$ is a cyclotron frequency
and we use new variables
\begin{align}
  & X := x^1 - \frac{p_2 + m~\calE/\calB}{g\calB}~,
&
  & E := \omega + \frac{1}{2m}~\frac{m\calE}{\calB}
       \left( \frac{m\calE}{\calB} + 2 p_2 \right)~.
\label{eq:def-X-calEpsilon}
\end{align}

The eigenvalue and eigenstate of Eq.(\ref{eq:reduced-EOMII}) are
\begin{align}
  & E_n = \left( n + \half \right) \omega_c~,
\label{eq:E-value} \\
  & \phi_n(x^1) = C_n~\exp\left( - \frac{m \omega_c}{2}~X^2 \right)~
       H_n\left( \sqrt{m\omega_c}~X \right)~,
\label{eq:E-state}
\end{align}
where $C_n$ is a constant and $H_n$ is the $n$-th Hermite polynomial
\begin{equation}
  H_n(z) := (-)^n e^{z^2}~\left( \frac{d}{dz} \right)^n~e^{-z^2}~.
\label{eq:def-Hn}
\end{equation}

For this solution $\displaystyle{
\psi_n(t, \vecx) =e^{-i \omega_n t}~\phi_n(x^1) \star e^{i p_2 x^2} }$,
we obtain
\begin{align}
  ( D_1 \star \psi_n ) \star \psi_n^\dagger
  &= \partial_1 \psi_n \star \psi_n^\dagger
   = \phi'_n (x^1) \star e^{i p_2 x^2} \star
                         e^{-i p_2 x^2} \star \phi^*_n(x^1)
\nonumber \\
  &= \phi'_n (x^1) \star \phi^*_n(x^1)
   = \phi'_n (x^1) \times \phi^*_n(x^1) \in \bmR
\label{eq:D1-psi-psi-dagger} \\
\nonumber \\
  ( D_2 \star \psi_n ) \star \psi_n^\dagger
  &= \partial_2 \psi_n \star \psi_n^\dagger
           - i~g\calB~x^1 \star \psi_n \star \psi_n^\dagger
\nonumber \\
  &= \phi_n (x^1) \star (i p_2)~e^{i p_2 x^2} \star
                                e^{-i p_2 x^2} \star \phi^*_n(x^1)
      - i~g\calB~x^1 \star \phi_n (x^1) \star e^{i p_2 x^2}
                            \star e^{-i p_2 x^2} \star \phi^*_n(x^1)
\nonumber \\
  &= i~( p_2 - g\calB~x^1 ) \star \phi_n (x^1) \star \phi^*_n(x^1)
\nonumber \\
  &= i~( p_2 - g\calB~x^1 )~\big\vert~\phi_n (x^1)~\big\vert^2~.
\label{eq:D2-psi-psi-dagger}
\end{align}
Therefore, the current (\ref{eq:def-J}) is expressed as
\begin{align}
  & J^0(t, \vecx)
   = g~\phi_n(x^1) \star \phi_n^*(x^1)
   = g~\big\vert~\phi_n(x^1)~\big\vert^2~,
\label{eq:J-zero} \\
  & J_1(t, \vecx)
      = 0~,
\label{eq:J-one} \\
  & J_2(t, \vecx)
  = \frac{1}{m}~( p_2 - g\calB~x^1 )~\big\vert~\phi_n (x^1)~\big\vert^2
  = - \rho_g(x^1) \left( \frac{\calE}{\calB}
          + \frac{g\calB}{m}~X \right)~,
\label{eq:J-two}
\end{align}
where $\rho_g := J^0$ is the charge density.

The NC $U(1)$ charge current {\it density} is gauge covariant,
but not {\it invariant},
while the total charge and current over space defined by
\begin{align}
  & Q := \int d^2x~J^0( t, \vecx)~,
&
  & \calJ_2 := \int d^2x~J_2( t, \vecx)~,
\label{eq:def-total-J-two}
\end{align}
respectively, are gauge invariant.
Since observable quantities should be gauge invariant,
we consider the total charge and total current.
For Eqs.(\ref{eq:J-zero}) and (\ref{eq:J-two}), we have
\begin{align}
  Q &= g~\int d^2x~\big\vert~\phi_n(x^1)~\big\vert^2~,
\label{eq:total-rhoE} \\
  \calJ_2 &= -g \int d^2x~\left( \frac{\calE}{\calB}
      + \frac{e\calB}{m}~X \right)~\big\vert~\phi_n (x^1)~\big\vert^2
     = -g~\frac{\calE}{\calB}~\int d^2x~
      \big\vert~\phi_n (x^1)~\big\vert^2
     = - \frac{Q}{\calB}~\calE~.  \label{eq:total-J-two}
\end{align}
Thus, between the total charge and total current,
we have the Hall conductivity
\begin{equation}
  \sigma_H = - \frac{Q}{\calB}~.
\label{eq:Hall-relation}
\end{equation}
The forms of Eqs.(\ref{eq:total-J-two}) and (\ref{eq:Hall-relation})
coincide with the usual commutative results,
except for the definition of $\calE$ and $\calB$.
Note that $\calB$ and $\calE$ are the field strengths of
NC $U(1)$ gauge field, not the usual $U(1)$.

We impose the periodic boundary condition with the periodicity
$(~L_1,~L_2~)$ for the wave function (\ref{eq:Separable}).
Hence, we have $p_2 L_2 = 2 \pi \hbar n = h n$, {\it i.e.},
$p_2=hn/L_2$.
Namely, each Landau level has a degeneracy.
The center of harmonic oscillator takes discrete values
\begin{equation}
  X_1 = \Delta X_1 \cdot n + \frac{m \calE}{g \calB^2}~,
\label{eq:def-Xone}
\end{equation}
where
\begin{equation}
  \Delta X_1 := \frac{h}{g \calB}~\frac{1}{L_2}~.
\label{eq:def-Delta-Xone}
\end{equation}
The number of the center inside the periodic box $(~L_1,~L_2~)$
is given by
\begin{equation}
  \frac{L_1}{\Delta X_1} = \frac{L_1 L_2}{S_0}~,
\end{equation}
where $S_0 := h/g\calB$ is considered as the effective area
occupied by one electron.
This means that the number of electron per unit area with same energy
is given by $1/S_0$.
The total magnetic flux penetrating this area is
\begin{equation}
  \phi_0 = \calB~S_0 = h/g~,
\label{eq:def-Dirac-flux}
\end{equation}
which corresponds to Dirac's quantized magnetic flux.

Now, let us define the filling factor by
\begin{equation}
  \nu := \frac{Q/g}{\calB/\phi_0}~.
\label{eq:def-filling}
\end{equation}
In terms of the filling factor the Hall conductivity is written as
\begin{equation}
  \sigma_H = - \nu~\frac{g^2}{h}~.
\label{eq:Hall-conduct-by-filling}
\end{equation}
If we take the composite model in the theory of quantum Hall effect,
the filling factor $\nu$ defined by Eq.(\ref{eq:def-filling}) can be
understood to represent the ratio of electron number
and quantized magnetic flux number where electrons and magnetic fluxes
make a bound state.
Hence the filling factor $\nu$ is expected generally to take
fractional numbers.
\section{concluding remarks}\label{sec:summary}
We have reconsidered the Hall effect on NC space.
In order to preserve the gauge symmetry of NC $U(1)$
we have worked with the classical NCFT and found that
the Hall conductivity on NC space does not depend on the NC parameter
$\theta$ and coincides with the usual commutative result, except for
the definitions of the field strength of the gauge field.

Here, we would like to make some comments on existing works.
Dayi and Jellal
\cite{ref:DayiJellal}
concluded
\footnote{There is a calculation error in Eq.(29) in their paper.}
that the Hall current on NC space is given by
\begin{align}
  & \langle~\hJ_y~\rangle = \sigma_H~E~,
&
  & \sigma_H = - \frac{e \rho}{B}~,
\label{eq:Hall-Dayi}
\end{align}
so that the Hall conductivity coincides with the conventional result.
However, their electric and magnetic fields are of the commutative
$U(1)$ gauge field, but not of the NC $U(1)$ gauge field, whereas ours
are of the NC $U(1)$ gauge field.
Their commutative $U(1)$ gauge field strengths $E$ and $B$ are
related to the NC $U(1)$ gauge field strengths $\calE$ and $\calB$ by
\begin{align}
  & \calE = \left( 1 - \theta~\frac{eB}{2} \right)~E~,
&
  & \calB = \left( 1 - \theta~\frac{eB}{4} \right)~B~.
\label{eq:NCEB-Dayi}
\end{align}
If they adopted NC $U(1)$ gauge field strengths,
they would obtain another form of the Hall conductivity
\begin{align}
  \sigma'_H
  = - \frac{ 1 - \theta~eB/4 }{ 1 - \theta~eB/2 }~
      \frac{e \rho}{\calB}~.
\label{eq:Hall-Dayi-NC}
\end{align}

In any way, two forms of the Hall conductivity,
Eqs.(\ref{eq:Hall-Dayi}) and (\ref{eq:Hall-Dayi-NC}), are
NC $U(1)$ gauge dependent,
in spite that the NC Schr\"odinger equation adopted by them
is NC $U(1)$ gauge covariant.
The reasons is that their definition of the charge current is not
gauge covariant.

The result by Duval and Horvathy
\cite{ref:DuvalHorvathy}
is same as ours.
In their work, however the gauge field and the Schr\"odinger equation
adopted by them are for the commutative $U(1)$ gauge group and hence
the NC $U(1)$ gauge invariance is not clear.

Kokado, et.al.\,%
\cite{ref:KokadoEtAL}
calculated the Hall conductivity in the case that
there exists noncommutativity not only among coordinates,
but also among momenta, by introducing the gauge field
with the same method as Duval and Horvathy's.
Although the system adopted by them is also primarily
for the commutative $U(1)$ gauge group,
the system accidentaly have the NC $U(1)$ gauge invariance
in the case that the field strength is constant.
Therefore, they used the definition for the NC $U(1)$ gauge field
as the field strength and obtained the Hall conductivity which
depends on the noncommutativity parameters.
However, it is not clear whether the system has
the NC $U(1)$ gauge invariance for general gauge field configurations.

Although, in order to preserve the gauge symmetry of NC $U(1)$
we have considered the Hall effect by using the classical NCFT,
our result is also valid for
quantized NCFT, provided that the Hamiltonian of the matter field
$\psi$ has bi-linear form and we treat the NC $U(1)$ gauge field
as an external background field.
Therefore, in order to find the $\theta$-dependence
in the Hall phenomena,
the implications of propagating mode of the NC $U(1)$ gauge field
or of scattering effect by impurities should be investigated.

As another approach to find the $\theta$-dependence, 
we may be able to make use of
the difference between the peak of particle number density
and that of the charge density.
The particle number current density $j$ is defined by
\begin{align}
  j^\mu(t, \vecx) :=
  \begin{cases}
     \psi^\dagger(t, \vecx) \star \psi(t, \vecx)
          & \mbox{~for~~} \mu = 0 \vspace{0.2cm} \\
   \displaystyle{
     \frac{1}{2\, m\, i}~\left[~
               \psi^\dagger \star ( D^\mu \star \psi )
           - ( D^\mu \star \psi )^\dagger \star \psi~\right] }
   = \frac{1}{m}~
      \Im \left[~\psi^\dagger(t, \vecx) \star
                       \big(~D_k \star \psi(t, \vecx)~\big)~\right]
                            & \mbox{~for~~} \mu \ne 0
  \end{cases}~,
\label{eq:def-j}
\end{align}
and is locally conserved $\displaystyle{ \partial_\mu j^\mu = 0 }$.
By the definition, the particle number density
\begin{equation}
  \rho(t, \vecx) := j^0(t, \vecx)
                  = \psi^\dagger(t, \vecx) \star \psi(t, \vecx)
\label{eq:def-rho}
\end{equation}
is gauge invariant.

For Eq.(\ref{eq:Separable}) this reduces to
\begin{align}
  \rho(t, \vecx)
  &= e^{-i p_2 x^2} \star \phi_n^*(x^1) \star \phi_n(x^1)
     \star e^{i p_2 x^2}
   = e^{-i p_2 x^2} \star \big\vert~\phi_n(x^1)~\big\vert^2
     \star e^{i p_2 x^2}
   = \big\vert~\phi_n(x^1 - \theta p_2 )~\big\vert^2~,
\label{eq:n-th-rho}
\end{align}
while the charge density $\rho_g$ for the same state is given by
Eq.(\ref{eq:J-zero}), so that $\displaystyle{%
\rho_g(t, \vecx) =g~\vert \, \phi_n(x^1) \, \vert^2 }$.

Although the particle number density cannot couple with
the NC $U(1)$ gauge field, it may couple with gravity.
On the other hand, the charge density can couple with the NC $U(1)$
gauge field.
By making use of these different properties,
it may be possible to provide an observable test for
noncommutativity of space.
The details of two things above will be considered elsewhere.

\end{document}